\documentclass[twocolumn]{revtex4}
\usepackage{graphicx}
\usepackage{scrtime}
\usepackage{latexsym}
\usepackage{epsfig}
\usepackage{fancyhdr}
\usepackage[hang]{subfigure}
\usepackage{supertabular}
\usepackage{wrapfig}
\usepackage{amssymb}

\begin{document}

\title{Vertical quantum wire realized with double cleaved-edge overgrowth}

\author{S. F. Roth}
\author{H. J. Krenner}
\author{D. Schuh}
\author{M. Bichler}
\author{M. Grayson}
\affiliation{Walter Schottky Institut, Technische Universit\"at M\"unchen, 85748, Garching, Germany}

\begin{abstract}
\noindent A quantum wire is fabricated on (001)-GaAs at the intersection of two overgrown cleaves.  The wire is contacted at each end to $n^+$-GaAs layers via two-dimensional (2D) leads. A sidegate controls the density of the wire revealing conductance quantization. The step height is strongly reduced from $2\frac{e^2}{h}$
due to the 2D-lead series resistance.  We characterize the 2D density and mobility for both cleave facets with four-point measurements.  The density on the first facet is modulated by the substrate potential, depleting a 2$\rm \mu m$ wide strip that defines the wire length. Micro-photoluminescence shows an extra peak consistent with 1D electron states at the corner. 
\end{abstract}

\maketitle

An elegant technique for fabricating clean one-dimensional (1D) quantum wires is cleaved-edge overgrowth (CEO) \cite{pfeiffer:1697}, where the 1D wire arises at the intersection of a 2D system and the perpendicular cleave plane. Ballistic transport \cite{yacoby:4612, piciotto:51}, optical spectroscopy \cite{hasen:54, Goni}, and lasing \cite{pfeiffer:laser, pfeiffer:laser2} in such single-cleave wires have been investigated.  Another advantage of CEO is that electron systems grown on the cleave facet can be modulated with atomic precision by the potential in the substrate.  2D CEO transport structures have been modulated with a superlattice potential \cite{deutschmann:1857}, and with single \cite{kang:59} and double tunnel barriers \cite{roth:2}.  The twin CEO capabilities of reducing the dimension and modulating the potential can be combined in double-cleave structures to create modulated 1D structures like quantum dot molecules \cite{wegscheider:1917, schedelbeck:1792}.  To date, however, no transport structures have been implemented by double-cleave due to the difficulty of fabricating electrical contacts.

This paper demonstrates a simple quantum wire in a double-cleave geometry with electrical contacts for transport measurements (Fig. 1). The wire develops at the intersection of two cleave planes in the accumulation edge of a depleted 2D electron system (2DES). Four-point measurements reveal the density and mobility of electrons in 
each overgrown facet \cite{finger},
and when a 2 $\mu$m wide low-density strip in the first cleave plane is depleted, the conductance shows quantization, indicating 1D transport at the corner.  The plateau steps are significantly reduced from $G_0=2\frac{e^2}{h}$ due to the 2D lead series resistance.
The quantum wire is also observed in photoluminescence as a second peak in the spectrum above the GaAs bandgap energy. 

\begin{figure}[!h]
\center
\includegraphics[width=8cm ,keepaspectratio]{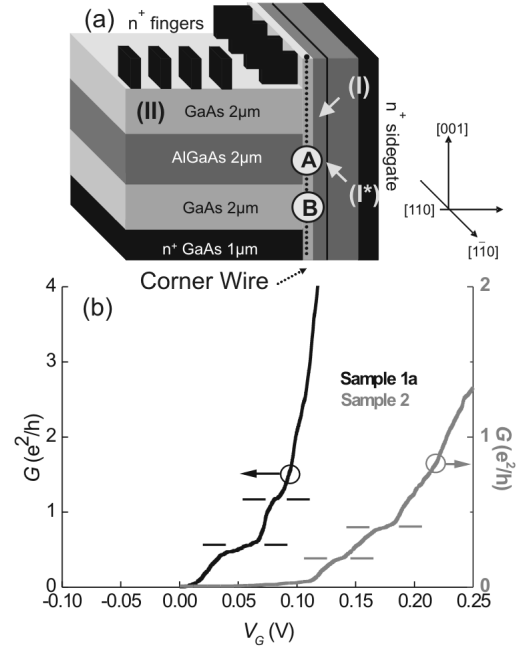}
\caption{(a) The double-cleave sample consists of an (001) grown substrate 
and a first $\rm (110)$ and second $\rm (1\bar{1}0)$ cleave, both overgrown with modulation doping. The 2DES regions from the two cleaves are indicated with (I) and (II), and the 2 $\mu$m substrate barrier defines a reduced density strip in the first cleave denoted as ($\rm I^*$).  The wire (dotted line) runs vertically at the corner of 2DES I and II and is contacted to the $n^+$ top- and bottom-contacts. The second CEO is in the plane of the front facet.
(b) The conductance of Sample 1a and 2 exposes steps between $V_G = \rm 0\:V$ and $\rm 0.2\:V$, indicative of 1D conductance quantization. The 2D lead resistance suppresses the step height from $2e^2/h$.}
\label{Sample}
\end{figure}

Figure 1a shows the design of the vertical quantum wire.  The substrate and first CEO are shown, with the second CEO taking place on the front facet. First, a (001) substrate is overgrown vertically with $\rm 1\:\mu m$ $n^+$ doped GaAs bottom-contact, $\rm 2\:\mu m$ undoped GaAs, $\rm 2\:\mu m$  $\rm Al_{0.3}Ga_{0.7}As$ (AlGaAs), another $\rm 2\:\mu m$ GaAs layer, and finally a $\rm 1\:\mu m$ $n^+$ top-contact.  This top most layer is patterned with photolithography and partly etched away, leaving eight $n^+$ finger-contacts \cite{finger}, which lead up to the cleave planes. 
Then the $\rm (110)$-facet is cleaved and overgrown giving rise to the first 2DES (2DES I). The overgrowth consists of an 8 nm wide GaAs quantum well, a 50 nm AlGaAs spacer, Si-$\delta$-doping, 500 nm AlGaAs, and a $n^+$-sidegate to control the 2D density.  The $2 \: \mu$m wide AlGaAs substrate barrier modulates the density of 2DES I, both due to quantum confinement \cite{deutschmann:1857} and because charged defects in AlGaAs can reduce the density next to the barrier \cite{poortere:153303, roth:1}. The reduced density region of 2DES I is notated 2DES I$^*$. To accumulate 1D states at the edge of 2DES I, the perpendicular $\rm (1\bar{1}0)$ cleave-facet is overgrown with a modulation doped barrier: 20 nm AlGaAs spacer, Si-$\delta$-doping and 300 nm AlGaAs topped with a 20 nm GaAs cap. The accumulation potential also induces a second 2DES (2DES II) at the second cleave plane, split by the substrate AlGaAs barrier.

Before measuring the wire, it is helpful to first characterize 2DES's I, $\rm I^*$ and II with four point magnetoresistance measurements using the $n^+$ finger-contacts. Standard lock-in techniques were used after illuminating the sample with a red LED.  Figure 2a and 2b show the Shubnikov-de-Haas oscillations measured at 300 mK in a magnetic field perpendicular to the first and second cleave facets, respectively.
 
At a sidegate bias $V_{\rm G}=0.0\:\rm V$ (inset Fig 2a) only one density 
is apparent in the resistance oscillations on the first facet.
The density and mobility $n_{\rm I}= 2.15\cdot 10^{11}\:\rm cm^{-2}$ 
and $\mu _{\rm I}=2.6\cdot 10^4\:\frac{\rm cm^2}{\rm Vs}$ are presumed to correspond to 2DES I, since 2DES $\rm I^*$ should be depleted.  When $n_{\rm I}$ is increased with sidegate bias to $V_{\rm G}=0.8\:\rm V$ (Fig 2a) the zero field resistance drops more than three orders of magnitude and additional minima appear. Both behaviours are evidence for the onset of conduction in 2DES I$^*$. The low resistance arises because of a short to the bulk-doped bottom-contact, and the additional minima arise from the lower density SdH-oscillations in 2DES I$^*$.  
The resistances $R_{\rm top-top}$ between two finger-contacts and $R_{\rm top-bottom}$ from a finger to the bottom-contact (Fig. 3 inset) confirm that two pinch-off conditions occur at $V_G$ = -0.1 V and +0.3 V, attributed to the depletions of 2DES I and 2DES $\rm I^*$, respectively. On the second cleave facet (Fig. 2b), the density and mobility of 2DES II is $n_{\rm II}=4.3\cdot 10^{11}\:\rm cm^{-2}$ and $\mu _{\rm II}=2.9 \cdot 10^4\:\frac{\rm cm^2}{\rm Vs}$, respectively.

\begin{figure}[!h]
\center
\includegraphics[width=7.5cm ,keepaspectratio]{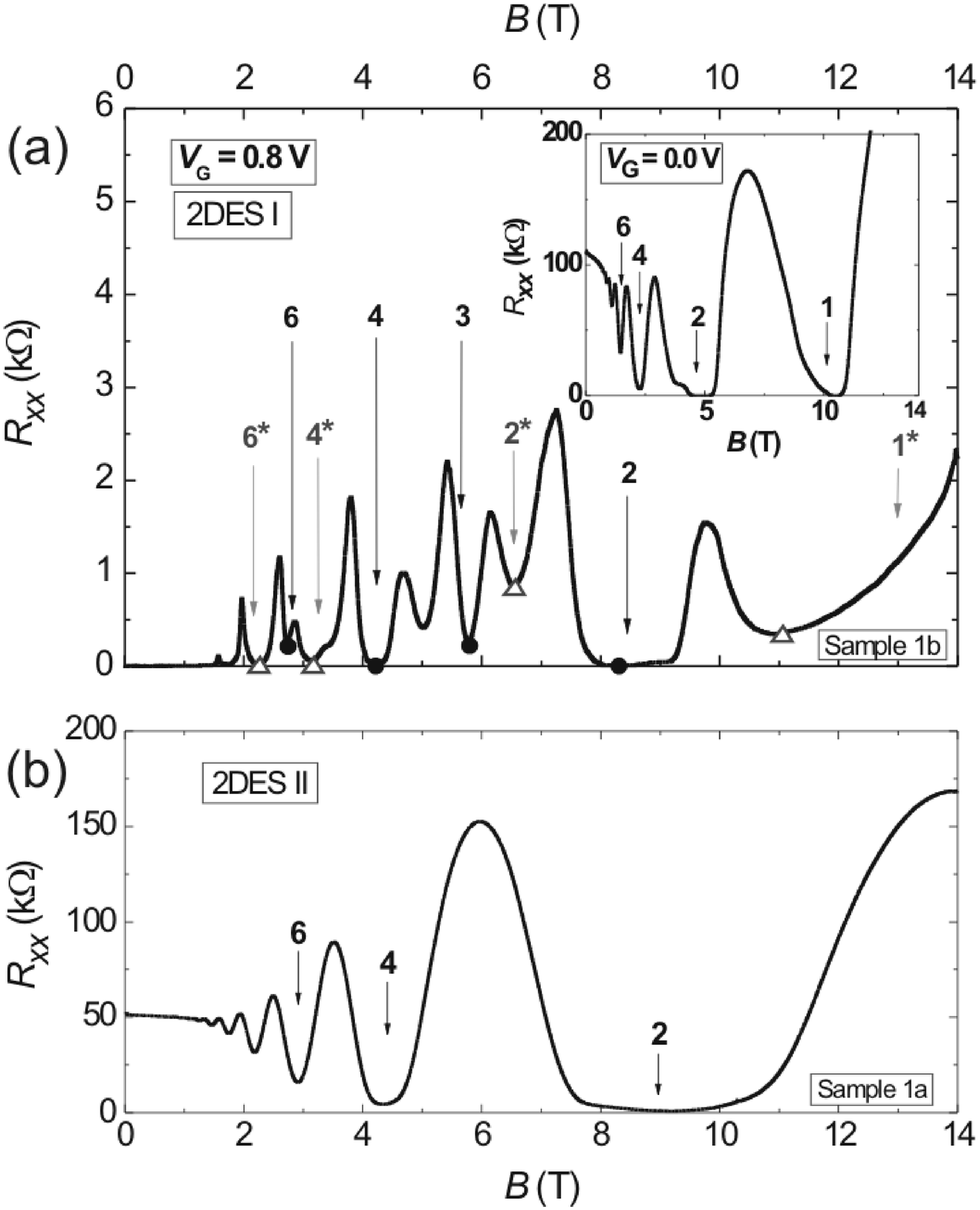}
\caption{Shubnikov-de Haas oscillations of 2DES I (a) for $V_{\rm G}=0.8\:\rm V$ and $V_{\rm G}=0.0\: \rm V$ (inset) and 2DES II (b). A sidegate tunes the density of 2DES I. Extra minima at $V_{\rm G}=0.8\:\rm V$ are attributed to a second density in region 2DES $\rm I^*$. Filling factors are labeled $\nu $ for density $n_{\rm I}$ of 2DES I and $\nu ^*$ for density $n_{\rm I^*}$ of 2DES $\rm I^*$. Symbols indicate minima in the data and arrows indicate the theoretical position of minima as described in the text.} 
\label{SDH}
\end{figure}

\begin{figure}[!h]
\center
\includegraphics[width=7.5cm ,keepaspectratio]{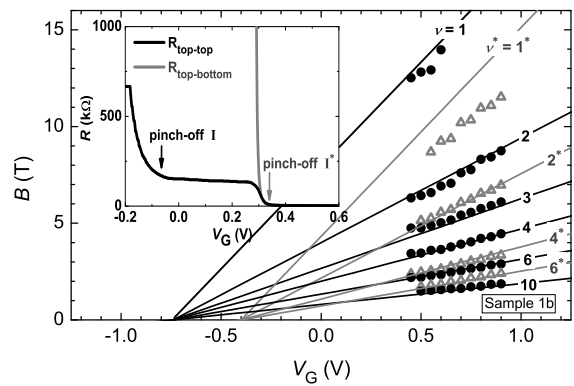}
\caption{Inset: The gate voltage dependent resistances $R_{\rm top-top}$ and $R_{\rm top-bottom}$ demonstrate the pinch-off of the low density central 2DES $\rm I^*$ and of the high density 2DES I with decreasing $V_{\rm G}$. The positions of the minima are plotted versus $V_{\rm G}$ and fit to two Landau fans. A density difference of $\Delta n= n_{\rm I} - n_{\rm I^*} = 0.92\cdot 10^{11}\:\rm cm^{-2}$ is able to explain all the observed minima.}
\label{landau}
\end{figure}

We can quantify the densities of 2DES I and $\rm I^*$ at positive gate voltage by plotting Landau fans of the $R_{xx}$ minima versus $V_{\rm G}$ (Fig.~3). The fit lines plot $B_\nu = \frac{n_{\rm I}\cdot h}{e\cdot \nu }$ (black) and $B_{\nu^*} = \frac{n_{\rm I^*}\cdot h}{e\cdot \nu^*}$ (gray), where $B_\nu$ is the magnetic field  of the $\nu^{\rm th}$ minima, and the densities $n_{\rm I}$ and $n_{\rm I^*}$ vary linearly with gate voltage with a fixed density difference $\Delta n = n_{\rm I} - n_{\rm I^*} = 0.92\cdot 10^{11}\:\rm cm^{-2}$.
The fits clearly identify all the observed minima.  Just as in the inset of Fig. 3, the pinch-off of 2DES I occurs -0.4 V relative to the pinch-off of 2DES $\rm I^*$, though 
differing cooldowns and illumination conditions shift the absolute gate voltage.  
At gate voltages where 2DES $\rm I ^*$ is depleted transport in the 1D accumulation edge of 2DES $\rm I ^*$ can be investigated.
 
Conductance steps as a function of sidegate bias in Fig.~1(b) are the principle evidence of 1D conduction in this structure. An ac-voltage of $\rm 100\: \mu V$ is applied between a top finger-contact and the bottom-contact, and the current is measured with lock-in techniques at $\rm 17 \: Hz$. In Fig. 1b the conductance of Sample 1a after illumination is zero at a gate voltage of $V_{\rm G}=0\:\rm V$, but rises to a plateau value of about $0.3\cdot G_0$ between $V_{\rm G}=0.03\:\rm V$ and $V_{\rm G}=0.07\:\rm V$. A second plateau appears at $0.6\cdot G_0$ between $V_{\rm G}=0.07\:\rm V$ and $V_{\rm G}=0.09\:\rm V$, indicating 1D transport. Sample 2 behaves similarly, with the conductance onset shifted to a slightly higher gate voltage and lower plateau values. The plateus in both samples are strongly suppressed from $G_0$, due to a high 2D lead resistance as well as possible 2D-1D scattering at the accumulation edge \cite{yacoby:4612, picciotto:1730}.   

\begin{figure}[!h]
\center
\includegraphics[width=\linewidth ,keepaspectratio]{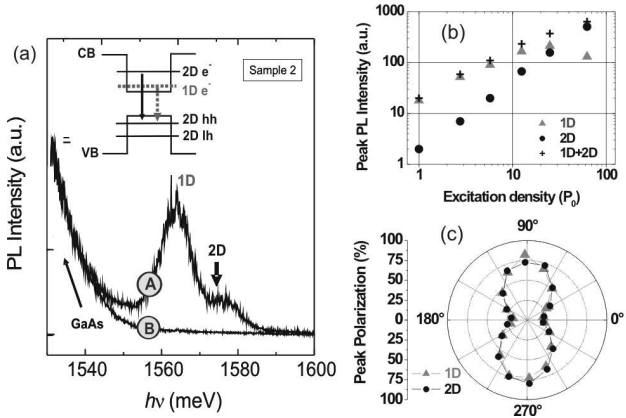}
\caption{(a) PL of a double-cleave sample measured on the square well (A) and the triangular well (B) (see Fig. 1). Two peaks appear at (A) and are attributed to a transition from 1D and 2D electron states to 2D heavy-hole states. The peaks disappear if the laser spot is moved to (B). The arrow indicates the calculated transition energy of the ground state. (b) At low laser powers only the 1D peak is visible. The 2D peak appears as the power is increased and electrons fill up higher states. When the lowest energy level is filled, the intensity of the 1D peak saturates. (c) Both peaks are strongly polarized as expected for transitions to heavy-hole states. The polarization angle $\theta=0^o$ is perpendicular to the quantum well growth direction.}
\label{Line}
\end{figure}

Evidence of a quantum wire is also observed optically via micro-photoluminescence ($\rm \mu$-PL). The sample is cooled down to 15 K in a flow-cryostat, and a HeNe laser ($\lambda =632.8\:\rm nm$) is focused to a diameter of $\rm 1\: \mu m$ on the $\rm (1\bar{1}0)$-facet (the front facet of Fig.~1a). Figure \ref{Line} shows PL spectra recorded for low excitation densities $(P_0=200\mathrm{\frac{\rm W}{\rm cm^{2}}})$. As the spot is postioned exactly above 2DES I$^*$ 
(position A in Fig. 1a), a large peak appears in the PL signal at a detection energy $h\nu=1562\: \rm meV$ and a smaller sidepeak at 1576 meV. When moving the laser spot along the [001] direction to rest above 2DES I, both peaks disappear (position B in Fig. 4a), because here no confinement for holes exists and only PL from the GaAs bulk is detected at the bandgap energy of 1514 meV. 
 
We calculate the interband absorption spectrum of a 8 nm GaAs quantum well embedded in a AlGaAs matrix. The lowest electron to heavy-hole transition was found to be at $E_{\rm e-hh}=1574\:\rm meV$, as indicated by the arrow in Fig. 4a. Accordingly, we attribute the high energy peak to a transition between 2D electron and 2D hole states. We expect to find the 1D electron states shifted to lower energy, as shown by the level diagram in Fig. 4a. Consequently, we assign the low energy peak to a transition between bound 1D electron and a continuum of 2D heavy-hole states \cite{Goni}. No holes are confined to 1D in our structure. The energy difference of 14 meV between the two peaks is the binding energy of the quantum wire and is in good agreement with Hartree calculations. 

A final test for the existence of a quantum wire is excitation power dependent and polarization resolved spectroscopy. The results are presented in Fig. 4b and 4c, respectively. For increasing excitation powers up to $100\cdot P_0$ both peaks gain intensity and the 1D peak shows a clear saturation behavior in contrast to the 2D signal. This observation can be explained by a consecutive filling of the 1D and 2D levels. Due to the higher density of states, the 2D peak gains intensity superlinearly, the 1D peak sublinearly. For the sum of both signals an exponent of $0.97\pm0.05$ was found, demonstrating the single exciton nature and the decay to a common hole level. In Fig. 4c we plot the degree of polarization $I(\theta)/(I_\perp+I_\parallel)$ of the 1D and 2D signal. For $\theta=0^o$ perpendicular to the quantum well growth direction of 2DES I, we find a maximum degree of polarization of $80\%$ for \emph{both} signals. Since the PL is detected perpendicular to the QW growth direction heavy-hole and light-hole states can be distinguished. In contrast to light-holes, heavy-hole transitions are polarized perpendicular to the QW growth direction due to the missing $z$-component of the central cell part of the wavefunction. The strong in-plane polarization of the 1D and 2D signal is a clear indication for a recombination of conduction band electrons with a heavy-hole state, ruling out a multiple peak structure from the light-hole energy spectrum. 
All four optical findings, the limited spatial area, the peak emission energy, the power dependent level filling and the heavy-hole polarization selection rules provide evidence for the formation of a bound 1D level for electrons below the 2D continuum.

In conclusion, we have fabricated a vertical quantum wire with twofold application of CEO. In conductance measurements, steps characteristic of 1D transport are observed. In optical investigations a peak in the PL spectrum on the low energy side of the quantum well signal is confirmed to arise from bound wire states by power and polarization dependent measurements. The vertical orientation of the wire makes it possible to fabricate samples with a 1D potential modulated by atomically precise barriers.  
\\
\\
\noindent This work is supported by the Bundesministerium for Bildung und Forschung (BmBF) through project 01BM469 and Deutsche Forschungsgemeinschaft Ab 35/4 and 35/5. We thank Jonathan J. Finley for assistance with PL measurements and Gerhard Abstreiter for discussions and continued support. \nocite{Sample:dcleave}

\end{document}